# Large Exotic Spin Torques in Antiferromagnetic Iron Rhodium

Jonathan Gibbons[1,2,3,4], Takaaki Dohi[5], Vivek P. Amin[6], Fei Xue[7,8], Haowen Ren[9,10], Jun-Wen Xu[9], Hanu Arava[11,3], Soho Shim[12,2], Hilal Saglam[13,3], Yuzi Liu[14], John E. Pearson[3], Nadya Mason[12,2], Amanda K. Petford-Long[11,3], Paul M. Haney[15], Mark D. Stiles[15], Eric E. Fullerton[10], Andrew D. Kent[9], Shunsuke Fukami[5], and Axel Hoffmann[1,2,12]

[1]Materials Science and Engineering, University of Illinois at Urbana-Champaign, Urbana, IL 61801, USA
[2]Materials Research Laboratory, University of Illinois at Urbana-Champaign, Urbana, IL 61801, USA
[3]Materials Science Division, Argonne National Laboratory, Lemont, IL 60439, USA
[4]Physics, University of California San Diego, La Jolla, CA 92093, USA
[5]Laboratory for Nanoelectronics and Spintronics, Research Institute of Electrical Communications, Tohoku University, Sendai 980-8577, Japan
[6]Physics, Indiana University - Purdue University Indianapolis, Indianapolis, IN 46202, USA
[7]Associate, Physical Measurement Laboratory, National Institute of Standards and Technology, Gaithersburg, MD 20899, USA
[8]Institute for Research in Electronics and Applied Physics & Maryland Nanocenter, University of Maryland, College Park, MD 20742, USA
[9]Center for Quantum Phenomena, New York University, New York, NY 10003, USA
[10]Center for Memory and Recording Research, University of California San Diego, La Jolla, CA 92093, USA
[11]Northwestern-Argonne Institute of Science and Engineering, Northwestern University, Evanston, IL 60208, USA
[12]Physics, University of Illinois at Urbana-Champaign, Urbana, IL 61801, USA
[13]Applied Physics, Yale University, New Haven, CT 06520, USA
[14]Center for Nanoscale Materials, Argonne National Laboratory, Lemont, IL 60439, USA
[15]Physical Measurement Laboratory, National Institute of Standards and Technology, Gaithersburg, MD 20899, USA

**ABSTRACT**

Spin torque is a promising tool for driving magnetization dynamics for novel computing technologies. These torques can be easily produced by spin-orbit effects, but for most conventional spin source materials, a high degree of crystal symmetry limits the geometry of the spin torques produced. Magnetic ordering is one way to reduce the symmetry of a material and allow exotic torques, and antiferromagnets are particularly promising because they are robust against external fields. We present spin torque ferromagnetic resonance measurements and second harmonic Hall measurements characterizing the spin torques in antiferromagnetic iron rhodium alloy. We report extremely large, strongly temperature-dependent exotic spin torques with a geometry apparently defined by the magnetic ordering direction. We find the spin torque efficiency of iron rhodium to be (330±150) % at 170 K and (91±32) % at room temperature. We support our conclusions with theoretical calculations showing how the antiferromagnetic ordering in iron rhodium gives rise to such exotic torques.

# I. INTRODUCTION

Spin orbit torques have drawn a large amount of attention as an efficient tool to electrically generate and transfer angular momentum to thin film magnetic systems. Such spin torques are generated when an applied charge current in a spin source material is converted into a spin current by spin-orbit coupling. The spin current then carries angular momentum into the magnetic system, exerting a torque on its magnetization [1,2,3,4]. For materials with sufficiently large spin-orbit coupling, such as β-W [5], Pt [6,7], and β-Ta [8], this technique can generate very large pure spin currents with low power cost and heat generation. The conversion efficiency between charge and spin current in a material is known as the spin torque efficiency. These spin-orbit torques may be used in conventional computing architectures to re-orient or switch nanomagnets [9,10,11] for non-volatile memory schemes [12,13], or in novel computing implementations to drive spin dynamics in more complex magnetic systems [14]. With a suitably efficient spin source material, these applications are poised to revolutionize low-power, high-speed computing strategies. Many of these applications are already being realized. For example, Grimaldi et al. have used this effect to achieve high-speed switching in heavy metal/nanomagnet devices [15]. Of particular note, Torrejon et al. and Romera et al. [16,17] have proven that spin torque-driven nanomagnetic oscillators can be effective implementations of neuromorphic computing, capable of improving speech recognition capabilities even for relatively simple implementations. Since spin-orbit torques may provide an even more energy efficient pathway for sustaining the necessary magnetization dynamics, one can expect that spin-orbit torque driven oscillators may be even better suited for large neuromorphic systems. Indeed, the control of spin torque oscillators via this effect is well-documented [18,19,20] and larger arrays of spin Hall nano-oscillators already show complex synchronization behavior that may be useful for neuromorphic computation [21]. For these and other applications, it is necessary to improve the efficiency of spin torque generation, either through larger spin orbit coupling or more effective spin torque geometries.

In high-symmetry materials, spin currents from spin-orbit coupling effects are geometrically constrained, such that the spin polarization of the charge current-induced spin accumulation points perpendicular to both the applied current and the spin current directions, allowing for only the small set of potential torques derived by Slonczewski [22] [Fig. 1 (a)]. The resultant torques are not very well suited for controlling perpendicularly magnetized nanomagnets or oscillators, since they are identical for both magnetization directions with respect to the surface normal. Recently, however, it has been shown that additional symmetry breaking, such as in lower-symmetry crystal structures, can allow the generation of spin torques of other forms, which we refer to as exotic, that can be better suited for practical applications [23]. Even in highly symmetric crystals, magnetic ordering can still break the underlying restrictive symmetries imposed by simple crystal structures and allow more complicated spin torques to be generated in the system [24,25,26,27,28,29]. Ferromagnetic materials, however, are susceptible to external magnetic fields and can be problematic due to magnetic coupling between the spin source layer and the target nanomagnet. Antiferromagnets, on the other hand, may still break the underlying symmetries, while being robust against applied fields. Such symmetry-breaking behavior has been observed, for instance, in the non-collinear antiferromagnets $IrMn_3$ and $Mn_3GaN$ [30,31,32]. It has also been predicted that the magnetic order in collinear antiferromagnets can be sufficient to reduce the symmetry of the system and allow for the existence of exotic torques [29,33]. In fact, any effect that creates a difference in spin lifetimes parallel to and perpendicular to the ordering direction can be expected to produce exotic torques [Fig. 1 (b)]. It is for instance possible that the localized exchange fields in an antiferromagnet may be sufficient to induce spin-decoherence of the component of spin perpendicular to the magnetic ordering direction, or in other words that the ordering may reduce the lifetime of spins not parallel to the magnetic order.

Iron rhodium stands out as a potentially excellent source of novel spin torques. It undergoes a temperature-driven transition from a low-temperature antiferromagnetic state to a high-temperature ferromagnetic state at around 350 K [34,35], but with a temperature varying based on growth conditions, stoichiometry, and applied magnetic field. Both of these states can be made stable near room temperature for practical device design. This magnetic ordering can be used to break the underlying crystal symmetry and allow for novel torques. In addition, the easy access to this near-room-temperature transition allows a comparison between the spin behavior in the two magnetic ordering states, and furthermore allows manipulation of the magnetic ordering by annealing through the transition. Beyond this, iron rhodium has been shown to have a reasonably large anomalous Hall effect, which may indicate large spin-orbit coupling [36]. Additionally, spin-dependent effects including the anomalous Hall effect have been shown to undergo temperature-driven transitions that could provide potentially interesting behavior [37]. In this paper, we present a study of the spin torque efficiency of FeRh in FeRh/Cu/$Ni_{80}Fe_{20}$ multilayers via complementary methods, including ordinary [7,8] and DC-biased spin torque ferromagnetic resonance [7] and the in-plane magnetized second harmonic Hall technique [38,39]. We demonstrate through these measurements the existence of a large non-conventional spin torque generation within FeRh. We further conduct temperature-dependent spin torque measurements to examine the role of antiferromagnetic ordering and band structure changes in

FeRh spin torque generation, demonstrating an enormous and highly temperature-dependent spin torque efficiency. Finally, we calculate the properties of the torques in FeRh and use this model to explain our results.

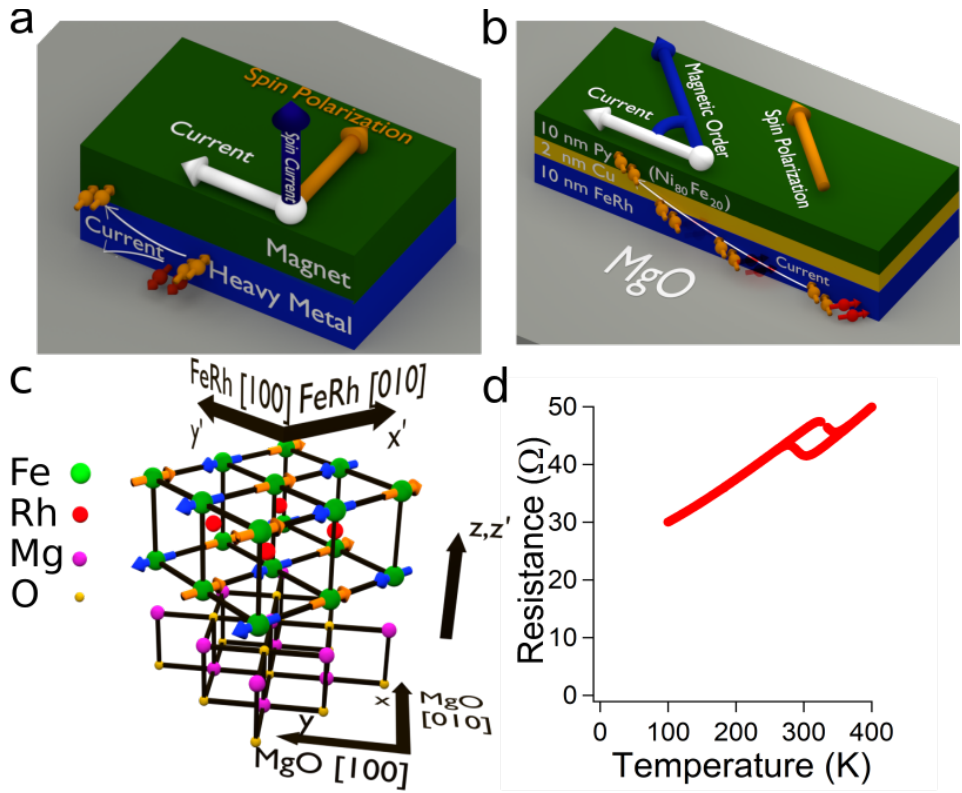

FIG. 1. (a) Ordinary spin Hall effect. Spin polarization is geometrically constrained to lie perpendicular to spin current and applied current. (b) Magnetic order in the FeRh breaks symmetries and allows more complex spin polarizations. (c) Atomic and magnetic structure of FeRh on MgO. FeRh order lies at a 45° angle with respect to the MgO cube directions. (d) Resistance measurements show the transition between antiferromagnetic and ferromagnetic states. For further transport measurements the sample is prepared in the antiferromagnetic state.

## II. SAMPLE STRUCTURE

We characterized the spin torque efficiency in devices using as the spin source material an epitaxial 50 % iron, 50 % rhodium (all compositions given in atomic fraction) alloy thin film grown on a (100) MgO substrate via sputtering. It is important to consider the crystal structure of the FeRh both for underlying symmetry considerations and for the effect of the crystal structure on magnetic anisotropy. MgO has been shown to cause strain effects that can lead to a preferred crystalline ordering direction in epitaxially-grown FeRh, with alternating planes of Fe and Rh along the film normal direction, and the cube axes at a 45° angle with respect to the underlying MgO cube orientation [40]. Moreover, these strain effects may even cause a tetragonal distortion that favors in-plane magnetic ordering. These strain effects thus likely favor a magnetic ordering direction in FeRh that lies in the film plane and at a 45° angle with respect to the MgO <100> crystal axes. The predicted crystal structure and magnetic structure of our sputtered FeRh is thus schematically shown in Fig. 1 (c).

We have prepared two different samples, Sample 1 and Sample 2. Sample 1 is composed of a 10 nm thick 50 % iron, 50 % rhodium ($Fe_{50}Rh_{50}$) alloy spin source layer, followed by a 2 nm thick copper spacer layer, a 10 nm thick permalloy (80 % nickel, 20 % iron) sensor layer, and finally a 3 nm thick aluminum capping layer. We begin the fabrication process by annealing magnesium oxide wafers in situ at 850 °C within our sputter system. We then grow FeRh at 450 °C via sputtering and perform another anneal at 650 °C to improve the film quality, relax the film stress and ensure a well-ordered crystal structure in our FeRh. Finally, we sputter the remaining films at room temperature. In order to characterize effects of the spacer layer including current shunting and spin decoherence, we also repeated this growth procedure for copper layer thicknesses ranging from 1 nm to 3 nm.

After this, we used photolithography to pattern our devices into 80 μm x 24 μm bars and etched them by ion milling. The devices are patterned parallel to one of the MgO crystal axes. During this process, we cooled the sample with liquid nitrogen to reduce film damage. These device proportions are chosen to provide a large vertical cross-section for spin travel and a sizable current density while maintaining a length several times the device width to remove edge effects. Finally, we used photolithography and lift-off techniques to fabricate titanium/gold waveguides to provide radio frequency contact to the patterned bars. The waveguides are patterned symmetrically to remove radio frequency (RF) artifacts caused by asymmetry. Using the same process, we also produced 120 μm x 20 μm Hall bars with lateral contacts for Hall voltage detection and 4-point resistance measurements. These devices are also patterned parallel to one MgO crystal axis.

Sample 2 is composed of the following layers: 20 nm $Fe_{50}Rh_{50}$, 4 nm copper, 5 nm permalloy, and 4 nm silicon oxide as a capping layer. This sample is pre-annealed for 20 min at 350 °C, then the FeRh film is grown at 350 °C and annealed for 45 min at 800 °C. Next, the other films are grown at room temperature. We then patterned and etched the samples into 60 μm x 10 μm and 18 μm x 3 μm bars and fabricated contact waveguides. These bars are again patterned parallel to one MgO crystal axis. Finally, Sample 2 was field cooled from 400 K with an applied magnetic field at a 45° angle to the devices (along a predicted FeRh crystal axis) to improve the likelihood that the magnetic ordering follows this crystal axis. These substantially different sample parameters allow us to ensure that our results are robust even for different thickness and growth conditions.

We confirmed the properties of our materials electrically and magnetically. Measuring the device resistance over our range of temperatures, we find that due to the hysteresis in the transition, our FeRh can be consistently prepared into either the antiferromagnetic or ferromagnetic phase at room temperature [Fig. 1 (d)]. From now on, we focus particularly on the antiferromagnetic phase.

### III. TRANSMISSION ELECTRON MICROSCOPY CHARACTERIZATION

Scanning transmission electron microscopy—energy-dispersive X-ray spectroscopy (STEM-EDS) and high-resolution (HRTEM) data provided us with the composition and morphology of the various layers and interfaces in the MgO/FeRh/Cu/$Ni_{80}Fe_{20}$/Al films. The Cu [Fig. 2 (e)] and Al [Fig. 2 (f)] EDS maps indicate low signals in the FeRh and $Ni_{80}Fe_{20}$ layers, however we believe that these very low signals are very likely originating from the sample holder— which is primarily made of Cu and Al. The sample was therefore found to be compositionally as expected, that is, each layer was compositionally homogenous, even though the interfaces are likely intermixed. The MgO/FeRh interface is intermixed across a region of 1 nm, across which the ratio of MgO to FeRh changes from 90 % to less than 10 % [see Supplementary Fig. 2 (b)]. However, it is much more difficult to quantify the width of the FeRh/Cu and Cu/Py interfaces, because of the contribution of the sample holder's signal to the Cu intensity [see Supplementary Fig. 2 (b)]. From the elemental maps, confirmed by the HRTEM images, we have measured the layer thicknesses to be as follows: FeRh = (10.3±0.3) nm, Cu = (3.2±0.1) nm, and $Ni_{80}Fe_{20}$ = (10.9±0.4) nm. HRTEM analysis confirms that the MgO substrate is a single crystal with a [001] surface normal. A fast Fourier transform (FFT) [Fig. 2 (h)] of a region in the MgO layer [Fig. 2 (j)] shows the (200) planes (d-spacing 0.21 nm), which we have used as an internal magnification calibration. The interface between the MgO and FeRh contains several misfit dislocations, and stacking faults are visible in the FeRh layer— an example of which is provided in Fig. 1 of the supplementary material. An FFT of a region in the FeRh layer [Fig. 2 (h)] shows that the FeRh is lattice matched along the [010] direction of the MgO, therefore the orientation relationship between the FeRh and the MgO is FeRh [110] (001) ∥ MgO [010] (002). The film normal of the FeRh is along the [110] direction when grown on MgO because FeRh has a B2 CsCl type structure with a lattice parameter of ≈0.29 nm, and thus the {110} planes have a spacing of ≈0.21 nm $(0.29/\sqrt{2})$, which matches the (200) plane spacing of MgO. The Cu and permalloy also grow epitaxially with an orientation relationship of [010] (002) ∥ MgO [010] (002).

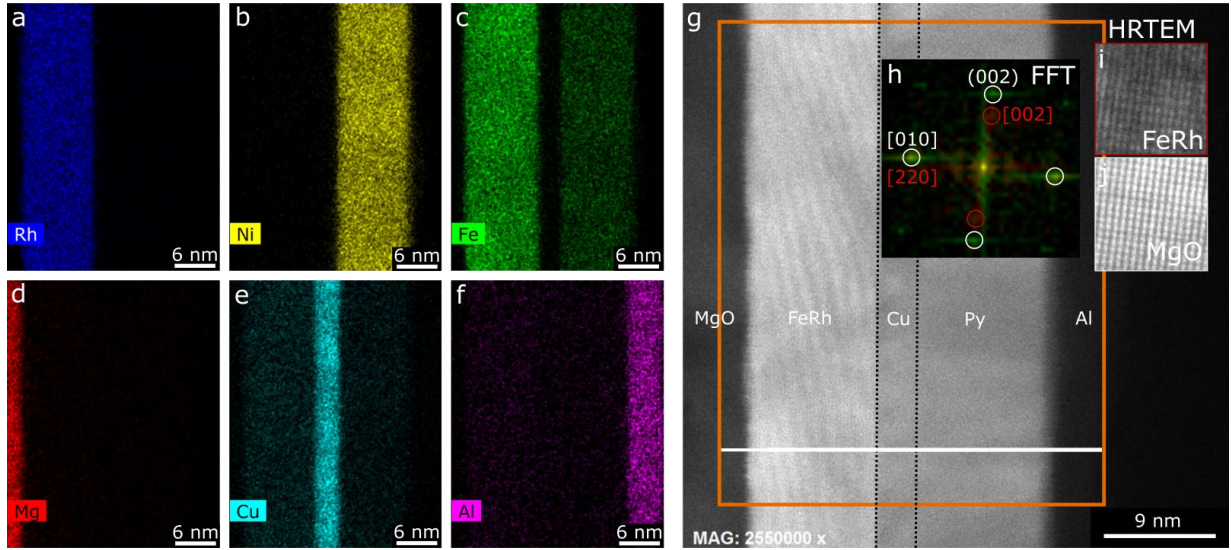

FIG. 2. EDS composition maps of a cross section of the MgO/FeRh/Cu/Py/Al multilayer for Rh (a), Ni (b), Fe (c), Mg (d), Cu (e), and Al (f). (g) High Angle Annular Dark Field (HAADF) STEM image of the multilayer cross section. (h) FFTs from the FeRh (red) and MgO (white) layers indicating lattice matching along the MgO [010] direction with a tetragonal distortion of the FeRh along the [001] direction of the MgO. (i–j) Magnified image of regions selected for FFTs: (i) FeRh and (j) MgO.

## IV. SPIN TORQUE FERROMAGNETIC RESONANCE LINE SHAPE ANALYSIS

We have used spin torque ferromagnetic resonance (ST-FMR) [7,8] to characterize the size of the transverse spin current generated in FeRh by an applied charge current [Fig. 3 (a)]. We performed our spin torque ferromagnetic resonance measurements under vacuum in a closed-cycle helium-cooled probe station using RF probes to contact our waveguides. The temperature on the station ranged from 300 K to below 170 K. We apply a 10 GHz RF current to our multilayer while an in-plane magnetic field saturates the permalloy layer at a direction, which is at a 45° angle with respect to the current direction.

The current flowing in the FeRh is converted into a spin current via spin-orbit effects and travels into the permalloy. The resulting spin torque drives the permalloy into resonance around the applied magnetic field direction. The precession of the permalloy modulates its resistance through anisotropic magnetoresistance (AMR), creating an RF-varying magnetoresistance which oscillates at the permalloy precession frequency. This varying magnetoresistance couples to the applied current to create a homodyne direct current (DC) mixing voltage, which is measured via a lock-in amplifier. By modifying the magnetic field strength, we control the permalloy precession frequency and sweep it through the applied current frequency, which allows us to measure the resonance line shape.

The resulting line shape is the sum of a symmetric and anti-symmetric Lorentzian, described by Eq. 1, and the strengths of the two components depend on the symmetry of the spin torques. Conventional spin Hall anti-damping torques (spin polarized in the y-direction) produce a symmetric line shape, while conventional spin Hall field-like torques produce an anti-symmetric line shape. The resulting signal may be modeled as:

$$V_{Mix} = S \frac{\Delta H}{(H_{ext}-H_0)^2 + \Delta H^2} + A \frac{\Delta H (H_{ext}-H_0)}{(H_{ext}-H_0)^2 + \Delta H^2}, \quad (1)$$

where $H_{ext}$ is the applied field, $H_0$ is the permalloy ferromagnetic resonance field, and $\Delta H$ is the resonance linewidth. $S$ and $A$ are fitting parameters designating the size of the symmetric and anti-symmetric Lorentzians, respectively. Typically, the dominant field-like torque in most materials is not a spin torque, but an Oersted torque. Oersted torques are well-understood, and therefore the anti-symmetric component of the line shape is typically used to calibrate the strength of a torque given the size of the resonance. Thus, by taking the ratio of the symmetric and anti-symmetric components of the line shape, we can estimate the size of the anti-damping spin torque and extract the spin torque efficiency. It is important to note here that only spins oriented perpendicular to the permalloy magnetization exert a torque on the permalloy and thus drive resonance. Thus, this technique only probes the component of the spin current with spin polarization perpendicular to the permalloy magnetization direction. We can thus define $\xi_\perp = \xi \sin(\phi_{ST})$ where $\phi_{ST}$ is the angle between the spin polarization and the permalloy magnetization. For conventional spin Hall torques, $\phi_{ST}$ is 45° and $\xi = \sqrt{2}\xi_\perp$. As we do not know the value of $\phi_{ST}$, we will use the shape analysis to extract the perpendicular component of the spin torque efficiency

$$\xi_\perp = \frac{J_s}{J_c} = \frac{S}{A} \frac{\sqrt{2}}{2} \frac{e\mu_0 M_s t d}{\hbar} \left(1 + \frac{4\pi M_{eff}}{H_{ext}}\right)^{\frac{1}{2}}, \quad (2)$$

where $t$ is the thickness of the permalloy layer, $d$ is the thickness of the iron rhodium layer, $M_s$ is the permalloy saturation magnetization and $M_{eff}$ is the effective magnetization of the permalloy for ferromagnetic resonance, which may be different from the saturation magnetization due to the effects of anisotropy. We assume that, due to the in-plane orientation of our permalloy, $M_s$ and $M_{eff}$ are equal.

We then repeat this measurement at various temperatures to characterize the temperature dependence of conventional spin Hall torques [Fig. 3 (b), (c)]. We find a reasonable value of (0.47±0.13) % for the perpendicular component of the spin torque efficiency at room temperature [Fig. 3 (d)], increasing to the higher value of (1.6±0.6) % at 170 K.

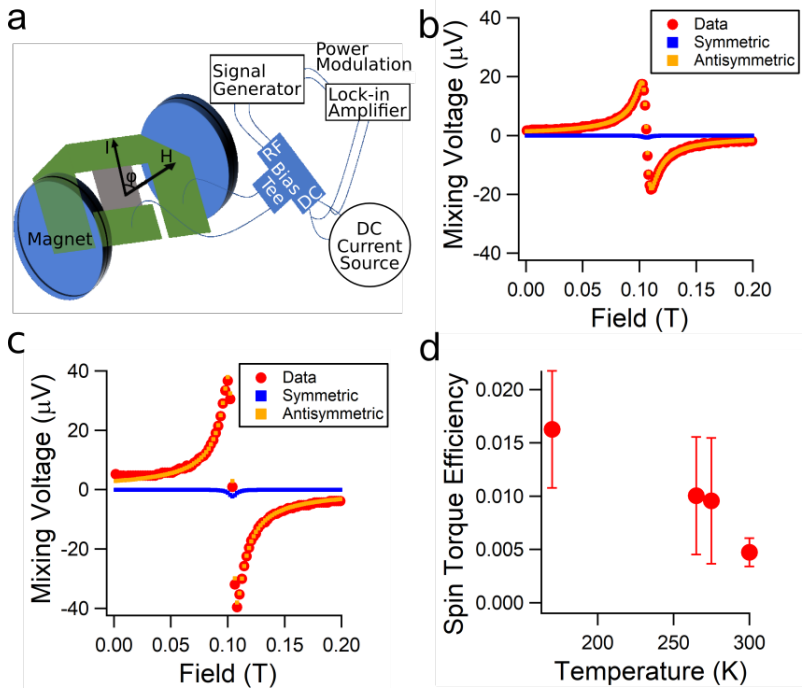

FIG. 3. (a) Diagram of our measurement setup for ST-FMR. RF current is applied through a bias tee by a signal generator, with the power modulated by a lock-in amplifier. The DC mixing voltage is measured through the bias tee by a lock-in amplifier. The magnetic field is applied at a 45° angle to the current direction. (b) A representative magnetic field scan of the resonance for Sample 1 at room temperature and (c) at 170 K, showing a very small symmetric component. (d) The perpendicular component of the spin torque efficiency extracted from line shape analysis. The spin torque efficiency is highly temperature dependent, increasing by a factor of ≈3.5 as the temperature is reduced. Error bars

indicate single standard deviation uncertainties extracted from the line shape analysis combined over a series of fitting conditions over multiple devices.

## V. DC-BIASED SPIN TORQUE FERROMAGNETIC RESONANCE

While the above ST-FMR method is well-suited to measuring conventional spin-orbit torques, it suffers from significant geometric restrictions. As mentioned above, if the spin current is polarized in the same direction as the magnetization, the magnet experiences no spin torque. This can be a problem for magnetically-ordered spin source layers if the magnetic order of the spin source and sensor layers are aligned parallel to each other. Signals from other sources, including heating, spin pumping [41,42] and various RF artifacts, can also interfere with the line shape analysis technique, leading to errors in the measured spin torque efficiency. Furthermore, the line shape analysis method described above assumes that the field-like torque in the system is sourced entirely by Oersted fields, an assumption which fails in materials with significant field-like torques or exotic torques. For these reasons, we require a more suitable characterization technique. Iihama et al. have used the DC-biased spin torque ferromagnetic resonance technique [7] to measure damping modulation due to spin currents produced in a layer with magnetic ordering [25]. The DC-biased spin torque ferromagnetic resonance technique characterizes the anti-damping torque without relying on assumptions about other torques in the system. Additionally, it probes the component of the spin current with spin polarization parallel to the magnetization, which makes it more appropriate for characterizing exotic spin torques generated from magnetic order.

We have thus used DC-biased spin-torque ferromagnetic resonance to confirm our observed temperature dependence and more accurately characterize the anti-damping torques generated in our FeRh. During ST-FMR measurements, we use a DC current source to apply a constant DC current to our device [Fig 3 (a)]. This creates a DC anti-damping spin torque that modifies the effective damping of the magnetic layer, which is reflected as a modification to the resonance linewidth. We measure the resonance over a range of applied DC bias currents [Fig. 4 (a), (b)] to detect the relation between damping modulation and applied DC current. The resulting data is fitted in order to extract the spin torque efficiency by using:

$$\Delta H = \Delta H_0 + \frac{\omega}{\gamma}\left[\alpha - \frac{\hbar \xi_\| J_c}{2e\left(H_{ext}+\frac{M_{eff}}{2}\right)M_s d}\right]. \quad (3)$$

Here, $J_c$ is the charge current density in the FeRh, α is the Gilbert damping coefficient, and ω is the applied current frequency. It is important to note that only the component of spin parallel to the magnetic ordering affects the damping, and thus this technique is sensitive only to the component of the spin current with spin polarization parallel to the sensor layer magnetization, $\xi_\parallel = \xi\cos(\phi_{ST})$, where $\phi_{ST}$ is once again the angle between the spin polarization and the permalloy magnetization. We must also take care to account for the reduction of signal due to current shunting in the permalloy sensor and copper spacer layers in calculating the value of $J_c$ (see Supplementary Information).

As a result, we see a very high spin torque efficiency for modulating the effective damping of our sensor magnet. We can thus conclude that the component of the spin current with spin polarized parallel to the permalloy magnetization is very large. For Sample 1, before accounting for current shunting, the spin torque efficiency is (24±6) % at room temperature, while at 170 K, it increases to around (85±33) %. After accounting for current shunting, the spin torque efficiency becomes (91±32) % at room temperature and (330±150) % at 170 K. For Sample 2, the spin torque efficiency is (21±8) % at room temperature, increasing to (33±13) % at 200 K. After accounting for current shunting, the spin torque efficiency becomes (35±14) % at room temperature and (57±23) % at 200 K. The efficiency is much larger than seen from the line shape analysis, but is similarly increased by a factor of approximately 3.5 when cooled from room temperature to 170 K. These measurements can be explained if the spin polarization of the spin currents generated by spin-orbit effects in the FeRh is nearly (but not quite) aligned parallel to the permalloy magnetization, in which case the small component of the spin polarization perpendicular to the magnetization gives rise to a small perpendicular component of the spin torque efficiency and thus generates a small symmetric component of the line shape. We know that there are likely small experimental angular misalignments in both patterning and magnetic field application, such that our magnetic field is not perfectly oriented with respect to the crystal structure and devices, which is expected to cause exactly such a misalignment.

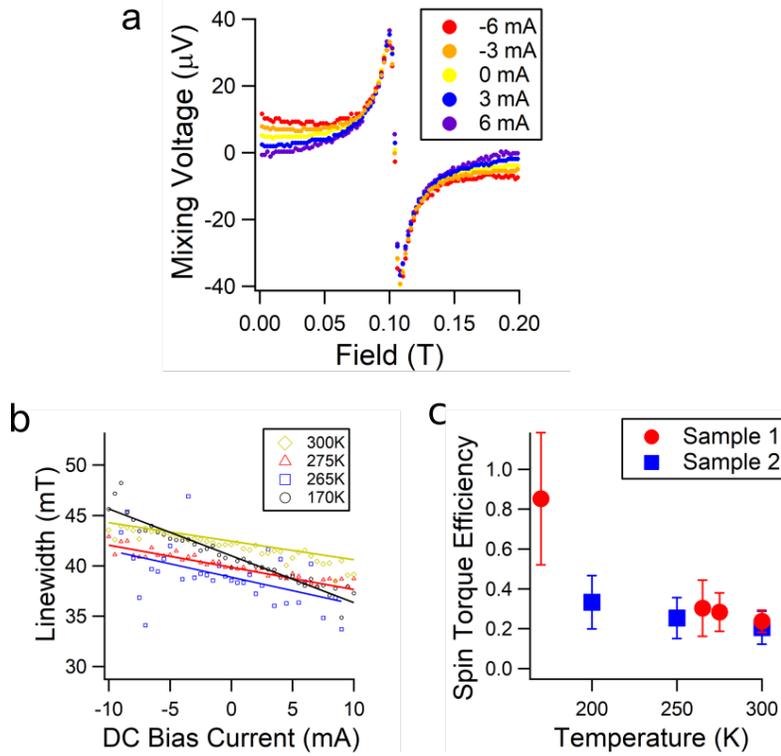

FIG. 4. (a) The ST-FMR resonance measured for different DC bias currents at 170 K for Sample 1. The linewidth is significantly modified at large currents. (b) The resonance linewidth over a range of DC bias currents for negative magnetic fields. Data is shown for 170 K, 265 K, 275 K, and 300 K. (c) The calculated spin torque efficiency from DC-biased ST-FMR. The efficiency increases by a factor of ≈3.5 between room temperature and 170 K, similar to the line shape analysis. Error bars indicate single standard deviation uncertainties extracted from the line shape analysis combined over a series of fitting conditions over multiple devices.

**VI. SECOND HARMONIC HALL MEASUREMENTS**

In order to further confirm our results and distinguish between ordinary spin Hall torques and exotic spin torques, we have performed additional measurements using the angle-dependent in-plane second harmonic Hall technique [38,39]. This technique is sensitive to spins polarized in any direction and can be used to help discern the true geometry of our generated torques. We perform these experiments on our Sample 1 Hall bar devices, composed as above of 10 nm FeRh, 2 nm Cu, 10 nm $Ni_{80}Fe_{20}$, and 3 nm Al. We apply a slowly varying sinusoidal DC current to a Hall bar and measure the second harmonic Hall voltage while varying the direction and strength of an in-plane magnetic field [Fig. 5 (a)]. Spin-orbit torques generated in the FeRh cause a rotation of the permalloy magnetization in a direction dependent on the symmetry of the torque. This rotation produces a change in the strength of the anomalous and planar Hall resistance of the permalloy, which then couples to the applied current to create a Hall resistance signal present on the second harmonic. This signal is measured as the in-plane field direction and strength of the applied magnetic field are varied. By fitting the field and angle dependence to a model based on the potential torque symmetries we obtain an expression for the second harmonic Hall voltage, $V_{2\omega}$:

$$V_{2\omega} = \frac{1}{2} I R_{PHE} \frac{H_{Oe}}{H} \cos(2\phi)\cos(\phi) + \frac{1}{2} I R_{PHE} \frac{H_{FLT}}{H} \cos(2\phi)\sin(\phi_s - \phi) + \frac{1}{2} I R_{AHE} \frac{H_{ADT}}{H+H_k} \sin(\phi_s - \phi), \quad (4)$$

where $\phi_s$ is the angle between the applied current direction and the spin polarization direction, $\phi$ is the angle between the current and the applied field direction, $I$ is the current, $H$ Is the applied field, $R_{PHE}$ is the planar Hall resistance, $R_{AHE}$ is the anomalous Hall resistance, $H_{Oe}$ is the Oersted field generated in the FeRh and Cu layers, $H_{FLT}$ is the effective field exerted by the field-like torque, $H_k$ is the anisotropy field of the magnet, and $H_{ADT}$ is the effective field exerted by the anti-damping torque. This enables us to separate the generated torques into their components and thus determine the spin polarization direction of the spin current generated in the FeRh. Our measured angular dependence for an example scan with 90 mT applied field at 300 K and 170 K is shown in Fig. 5 (b) and (c).

We determine the direction of the spin polarization to be neither parallel to nor perpendicular to the applied current direction, confirming that it is an exotic torque. We observe both an anti-damping torque and a field-like torque with angular dependence indicating such a spin polarization. Although thermal effects (see Supplementary Information) and Oersted-driven field-like torques prevent us from determining the exact spin polarization angle, it is evidently somewhere close to a 45° angle. Unfortunately, the same parasitic thermal signals make it difficult to quantify the exact size of the total torque but based on the size of the exotic torque alone, we can estimate the size of the total torque assuming a 45° polarization. Using this method of estimation, we find a spin torque efficiency of about (16±6) % at room temperature and (41±13) % at 170 K, which is consistent with the values measured from ST-FMR. After accounting for current shunting, these efficiencies become (62±28) % at room temperature and (158±63) % at 170 K. We also estimate the size of the spin torque efficiency for the field-like torque in the same manner to be about (7±3) % at room temperature and (14±5) % at 170 K, or (27±12) % at room temperature and (54±23) % at 170 K after accounting for current shunting.

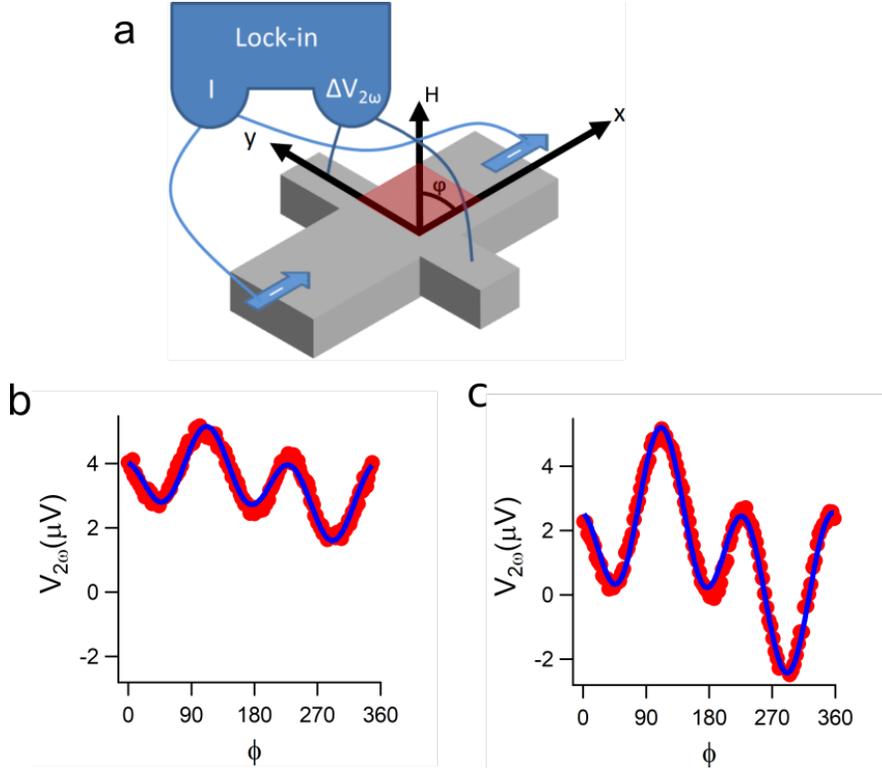

FIG. 5. (a) Schematic diagram of second harmonic Hall measurement setup. (b, c) Second harmonic Hall data with fit as magnetic field is rotated in the sample plane (b) at room temperature and (c) at 170 K.

## VII. DISCUSSION: GEOMETRY AND TORQUES

When performing ST-FMR in heavy metal/ferromagnet bilayers, the spin Hall effect generates a spin current polarized perpendicular to the applied current direction, whereas the magnetization of the ferromagnet is generally saturated at a 45° angle with respect to the current. Thus, the line shape analysis and DC bias current dependence of the ST-FMR technique are expected to yield the same value for the spin torque efficiency, because the spin polarization lies at a 45° angle with respect to the magnetization. However, in general, the size of the two effects depends on the direction of the spin polarization. As has already been mentioned, magnetic ordering reduces the symmetry of the material and allows for ordinarily forbidden spin polarization directions that yield different expected values for the two techniques. We conclude from the small size of the symmetric line shape contribution and the large modification to the damping, that the majority of the spin current has a spin polarization parallel to the applied magnetic field direction and thus parallel to the permalloy magnetization. In this case, the resulting RF torque does not drive the magnet into resonance, but the DC torque does modify the effective damping of the permalloy. Thus, we can conclude that the spin polarization of the spin current generated in the FeRh aligns itself at a 45° angle with respect to the fabricated devices.

In the simplest cases, such as the anomalous spin Hall effect, the spin polarization of the spin current preferentially aligns itself with the magnetic order of the spin source material, either parallel to or perpendicular to the order. For our FeRh under magnetic fields less than 1 T, the magnetic field does not define the Néel order direction. Instead, as discussed earlier, the ordering is most likely defined by material parameters, preferentially aligning with the crystal axes of the material, at a 45° angle with respect to the MgO crystal structure. Our devices are patterned parallel to the MgO substrate crystal axes, and thus the FeRh Néel order likely lies at a 45° angle with respect to our applied current, which is either parallel to or perpendicular to our measured spin polarization direction. For Sample 2, we performed field cooling to define the Néel order and saw similar results. We thus conclude that our FeRh generates an exotic spin current with spin polarization either parallel or perpendicular to its Néel order. This conclusion is further supported by our second harmonic Hall data, which shows that the spin polarization lies at some intermediary angle close to 45°. As a result, we can conclude that the spin polarization of the spin current generated in FeRh is most likely defined by its Néel order.

## VIII. THEORETICAL CALCULATION OF THE INTRINSIC SPIN CURRENT CONDUCTIVITY OF FERH

To compare the experimental results with standard theoretical approaches, we compute the spin current generated in FeRh via the intrinsic mechanism [43]. When the electric field is parallel or perpendicular to the Néel vector, the system symmetry constrains the spin Hall conductivity to assume a conventional form in which the electric field, spin current flow and spin polarization directions are mutually orthogonal. However, the spin current conductivity is anisotropic: the transversely flowing spin current is different if the electric field is parallel versus perpendicular to the Néel vector. We define $\sigma_\parallel$ ($\sigma_\perp$) to be the spin current conductivity when the spin polarization of the spin current is parallel (perpendicular) to the Néel vector. (Note that $\sigma_\parallel$ ($\sigma_\perp$) applies when the electric field is perpendicular (parallel) to the Néel vector.) We choose a sign convention in which a positive value of $\sigma_{\parallel,\perp}$ indicates that the field, flow, and spin directions form a right-handed coordinate system.

$$\vec{s} = \frac{\sin(2\phi)}{2}(\sigma_\parallel - \sigma_\perp)\hat{x} + (\sigma_\parallel \sin^2\phi + \sigma_\perp \cos^2\phi)\hat{y}. \tag{5}$$

To compute $\sigma_\parallel$ and $\sigma_\perp$, we use Quantum ESPRESSO [44] and Wannier90 [45]. FeRh has a cubic structure with unit cell length $a = 0.3019$ nm. In the Quantum ESPRESSO implementation, we use the Optimized Norm-Conserving Vanderbilt pseudopotentials [46] generated with a fully relativistic calculation using Perdew-Burke-Ernzerhof exchange correlations [47]. An $8 \times 8 \times 8$ Monkhorst-Pack mesh [48] is used with a 1000 eV cutoff energy. We then use Wannier90 [45] to obtain the Hamiltonian in an atomic basis, projecting the plane-wave solutions onto atomic $s, d$ orbitals of the Fe and Rh atoms. To calculate the spin current response, we utilize the Wannier interpolation within the tight-binding approximation, leading to the following form of the Kubo formula [49]:

$$\sigma_{ijk} = -2\frac{e^2}{\hbar} I \sum_k \sum_{n \neq m} \frac{\langle n|Q_{ij}|m\rangle\langle m|v_k|n\rangle}{(\varepsilon_n - \varepsilon_m)^2 + \eta^2}, \tag{6}$$

where $i, j, k \in [x', y', z']$ denote directions, $n, m$ label eigenstates at a given $k$-vector, $\varepsilon_{n,m}$ are the corresponding energies, $v_k$ is the velocity operator along the $k$-th direction, $Q_{ij} = (v_i \otimes s_j + s_j \otimes v_i)/2$ is the spin current operator, $s_j$ is the $j$-th Pauli matrix, and $\eta = 25$ meV is the broadening energy. Note that using this notation, $\sigma_\parallel = \sigma_{z'x'y'}$ and $\sigma_\perp = \sigma_{z'y'x'}$. Fig. 6 (b) shows the results of this calculation over a range of Fermi energies near the Fermi level to reflect general trends.

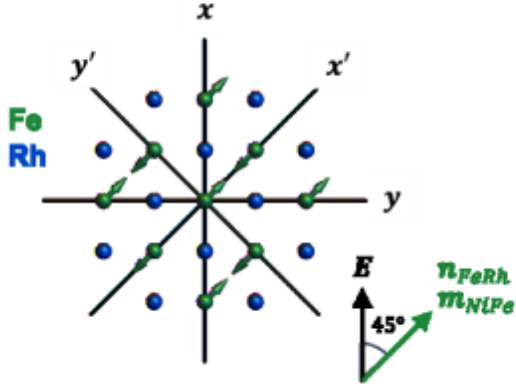 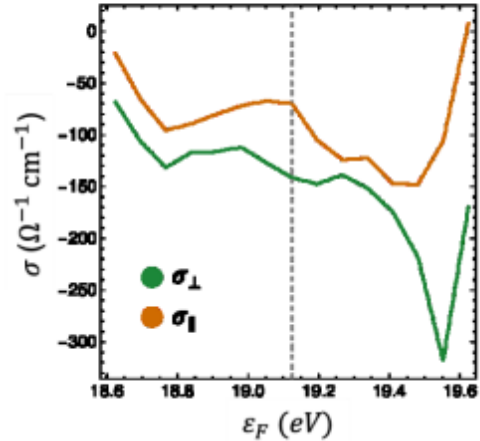

FIG. 6. (a) System geometry. The coordinate system $(x', y', z')$ describes the high symmetry directions of FeRh while $(x, y, z)$ describes the high symmetry directions of the MgO substrate, where $x' = \frac{1}{\sqrt{2}}(x + y)$, $y' = \frac{1}{\sqrt{2}}(x - y)$, and $z' = z$. (b) Theoretical calculations of the intrinsic spin current in FeRh under an applied electric field $\vec{E} \parallel \hat{x}$. The plot shows the spin current conductivity versus Fermi energy

for the parallel ($\sigma_\parallel$) and transverse ($\sigma_\perp$) spin conductivities defined above. The dashed line denotes the calculated Fermi energy $\varepsilon_F = 19.124$ eV at zero temperature.

The theoretical calculations show that for an applied, in-plane electric field oriented at a 45° angle from the Néel order parameter, the out-of-plane flowing spin current has a component of spin direction both parallel and transverse to the Néel order. In domains in which the electric field is 45° from the Néel order parameter, the spin conductivity for spins parallel to the Néel order is $\sigma_\parallel^* = \sigma_\parallel/\sqrt{2} = 48.4$ $\Omega^{-1}$cm$^{-1}$ and the spin conductivity for spins transverse to the Néel order is $\sigma_\perp^* = \sigma_\perp/\sqrt{2} = 99.7$ $\Omega^{-1}$cm$^{-1}$. In domains in which the electric field is -45° from the Néel order parameter, the spin conductivities are the same but are now defined at 90° from those in the first set of domains. If both domains are present, the two conductivities average to 74.1 $\Omega^{-1}$cm$^{-1}$, characteristic of a typical spin Hall effect. These calculations show that the two spin current components are sensitive to the Néel vector orientation and can be different, as seen in experiment. The differences are smaller than are seen in experiment but the variation of the conductivities with the Fermi level show that the similarity of two components is not a general conclusion. There are Fermi levels at which the ratio of the two is much larger corresponding to what is seen experimentally. Note however, that the two values being different, as measured, requires that one antiferromagnetic domain is more prevalent than the other.

## IX. CONCLUSIONS

We have conducted spin torque ferromagnetic resonance measurements to investigate the effect of magnetic ordering on the generation of spin currents in collinear antiferromagnetic iron rhodium, and have measured the resulting spin torques to be large and highly temperature dependent, with a spin torque efficiency of (91±32) % at room temperature and (330±150) % at 170 K. Further, we have shown the generated spin currents have a spin polarization defined by the magnetic ordering of the material, giving rise to spin torques with exotic geometries. We have also confirmed both the large spin torque efficiency and the exotic nature of the spin torques via second harmonic Hall measurements. Lastly, we have also carried out theoretical band structure calculations that support our results, indicating that the magnetic ordering in FeRh creates asymmetric spin conductivities consistent with our measured results. These results suggest that it is likely possible to generate large, magnetically controllable, exotic spin torques in FeRh. Such customizable exotic torques have a great deal of utility for practical spintronics device design, and may be of particular use for neuromorphic computing via spin torque nano-oscillators and magnetic memory via nanomagnet switching.

## ACKNOWLEDGMENTS

We would like to thank I. Schuller, S. Siddiqui, M. Lonsky, M. Vogel, and Y. Li for useful discussions relevant to this research. This work was supported as part of the Quantum Materials for Energy Efficient Neuromorphic Computing, an Energy Frontier Research Center funded by the U. S. Department of Energy, Office of Science. The second harmonic Hall measurements were also supported by the NSF through the University of Illinois at Urbana-Champaign Materials Research Science and Engineering Center under Grant No. DMR-1720633 and was, in part, carried out in the Materials Research Laboratory Central Research Facilities, University of Illinois. Use of the Center for Nanoscale Materials, an Office of Science user facility, was supported by the U.S. Department of Energy, Office of Science, Office of Basic Energy Sciences, under Contract No. DE-AC02-06CH11357. The theoretical calculations in this work were partly supported by the National Institute of Standards and Technology, U. S. Department of Commerce. A portion of this work was also supported by the JSPS Kakenhi No. 19H05622. T.D. acknowledges the financial support from the GP-Spin of Tohoku University.